\documentclass[12pt]{article}
\usepackage[dvips]{graphicx}
\usepackage{latexsym}
\usepackage{amssymb}

\newcommand{\C}{\mathbb{C}}
\newcommand{\CP}{\mathbb{CP}}

\newcommand{\R}{\mathbb{R}}
\newcommand{\Z}{\mathbb{Z}}

\renewcommand{\d}{\mathrm{d}}

\newcommand{\koniec}{\begin{flushright}  $\Box $ \end{flushright}}
\def\be{\begin{equation}}
\def\ee{\end{equation}}

\def\Sm{\Sigma}

\def\Om{\Omega}

\def\om{\omega}

\def\p{\partial}

\def\sm{\sigma}

\newcommand{\hook}{{\setlength{\unitlength}{11pt}   
                   \begin{picture}(.833,.8)
                   \put(.15,.08){\line(1,0){.35}}
                   \put(.5,.08){\line(0,1){.5}}
                   \end{picture}}}
\def\a{\alpha}
\def\l{\lambda}

\newtheorem{theo}{Theorem}[section] 
\newtheorem{prop}[theo]{Proposition}  

\newtheorem{defi}[theo]{Definition}
\newtheorem{col}[theo]{Corollary}

\begin{document}
\title{Einstein--Weyl spaces and dispersionless 
Kadomtsev--Petviashvili equation from Painlev\'e I and II.}
\author{Maciej  Dunajski, Paul Tod
\\ The Mathematical Institute,
24-29 St Giles, Oxford OX1 3LB, UK}  
\date{} 
\maketitle
\abstract {We present two constructions of new solutions to
the dispersionless KP (dKP) equation  
arising from the first two Painlev\'e transcendents.
The first construction is a hodograph transformation based on
Einstein--Weyl geometry, the generalised Nahm's equation and 
the isomonodromy problem.
The second construction, motivated by the first, 
is a direct characterisation of 
solutions to  dKP which are constant on a central
quadric.

We show how the solutions to the dKP equations can be used to construct
some three-dimensional Einstein--Weyl structures, and four--dimensional
anti-self-dual null-K\"ahler metrics.
} 
\noindent
\section{Introduction}
\label{INTR}
Let ${\cal W}$ be a three-dimensional complex manifold,
with a torsion-free connection
$D$ and a conformal metric $[h]$.
We shall call  ${\cal W}$ a Weyl space if
the null geodesics of $[h]$ are also geodesics for $D$. 
This condition is equivalent to 
$
Dh=\om\otimes h
$
for some one form $\om$. Here $h$ 
is a representative metric in
the conformal class.
If we change this representative by
$h\longrightarrow \phi^2 h$, then $\om\longrightarrow
\om+2\d\ln{\phi}$.  
A tensor object $T$ which transforms as $
T\longrightarrow \phi^m T$ when
$ h\longrightarrow \phi^2 h$
is said to be conformally invariant of weight $m$.
The pair $([h], D)$ satisfies the Einstein--Weyl (EW) equations if
the symmetrised Ricci tensor of of the Weyl connection is proportional
to the conformal metric \cite{C43,H82}.

In \cite{DMT00} it has been demonstrated that
if an Einstein--Weyl space  admits
a parallel weighted vector, then coordinates can be found in which 
the metric and the one-form are locally given by 
\be
\label{EWdkp}
h=\d y^2-4\d x\d t-4u\d t^2,\qquad\om=-4u_{x}\d t,\qquad u=u(x, y, t)
\ee
and the Einstein--Weyl 
equations reduce to 
the dispersionless Kadomtsev--Petviashvili equation
\be
\label{dkp}
(u_t-uu_x)_x=u_{yy}.
\ee
If $u(x, y, t)$ is a smooth real function of real variables then
(\ref{EWdkp}) has signature $(++-)$.
One can verify that the  vector $\p_x$ in the EW space (\ref{EWdkp})
is a null vector, covariantly constant in the Weyl connection, and  
with weight $-1/2$.

The geometric approach based on the structure (\ref{EWdkp}), and the
associated twistor theory yielded some new explicit solutions to
(\ref{dkp})\cite{DMT00}. 
Other solutions have been obtained in \cite{GK89,KMR01}.

In this paper we aim to show a nontrivial  relation between EW geometry
and the dKP equation on one side and the first two  Painlev\'e
transcendents on the other. 
The non-triviality here means that the Painlev\'e equations  
do not arise as symmetry reductions of dKP.

Let $\ast:\Lambda^i({\cal W})\rightarrow \Lambda^{3-i}({\cal W}), i=1, 2, 3$ 
be the Hodge operator corresponding to $[h]$.
Using the relations
\[
\ast\d t=\d t\wedge\d y,\qquad \ast\d y=2\d t\wedge\d x,\qquad 
\ast\d x=\d y\wedge\d x+2u\d y\wedge\d t
\]
we verify that equation (\ref{dkp}) is equivalent to 
\be
\label{gdkp}
\d\ast \d u=0.
\ee 
One needs to consider two classes of solutions:
\begin{enumerate}
\item 
The generic case $|\d u|^2=\d u\wedge\ast\d u\neq 0$. The condition
(\ref{gdkp}) implies the existence of a foliation ${\cal W}=\C\times
\Sm$ of an EW space (\ref{EWdkp}) by two-complex-dimensional symplectic manifolds
$\Sm$ with a holomorphic symplectic two-form $\om$ such that
\be
\label{poincare}
\ast\d u=\om.
\ee
We shall consider two cases: $\Sm=\C^2, \om=\d p\wedge\d q$, and
$\Sm=\CP^1\times\CP^1, \om=(1+pq)^{-2}\d p \wedge\d q$. The functions 
$p, q:\cal W\longrightarrow \C$ are local holomorphic
coordinates on $\Sm$, and in the second case the symplectic two-form
is defined on the complement of $1+pq=0$. 

The idea is to use
$(u, p, q)$ as local coordinates, and regard
$
x=x(u, p, q), y=y(u, p, q), t=t(u, p, q)
$
as dependent variables.
\item
The null case $|\d u|=0$. The 
existence of smooth $p$ and $q$ can not be deduced
even locally.
\end{enumerate}
In the next Section we shall see that Case (2) 
can be integrated completely (Proposition \ref{constraint}).
In Section \ref{Nahm} we shall show that Case (1) leads to a hodograph transformation between solutions to (\ref{dkp}) and solutions to the generalised
$\mbox{SDiff}({\Sm})$ (or so called $SU(\infty)$) 
Nahm equation. The special case of $SL(2,\C)$
generalised Nahm's equations turns out to be equivalent to the
isomonodromy problem associated with the Painlev\'e I  or Painlev\'e II equations.
The level sets of corresponding solutions to the dKP equations are 
central quadrics in the $(x, y, t)$ space (Theorem \ref{mainth}).
This will be established in Section \ref{Spainleve}.
In Section \ref{quadricAN} we shall discuss a converse construction:
if a solution to  dKP is constant on a central
quadric in the sense that
$u(x, y, t)$ is determined implicitly by an equation of the form
\be
\label{quadric}
Q(x, y, t, u)={X}^T{\bf M}(u){X}=C,
\ee
where $C$ is a constant, $X^T=(x, y, t)$, 
and ${\bf M}(u)$ is a symmetric matrix whose
components depend on $u$, then it 
is determined by one of the first two Painlev\'e equations
(Theorem \ref{Qquadric}).
Rational solutions to Painlev\'e II yield explicit new solutions to 
(\ref{dkp}). Examples of these are worked out in Section \ref{Sexamples}.
Finally in Section \ref{Sgeometries} we relate the Painlev\'e solutions
of dKP to Einstein--Weyl and anti-self-dual null-K\"ahler geometries.
\subsection{Users guide}
Although the representation (\ref{poincare}) which underlies the
results of this paper has its roots in the EW structure (\ref{EWdkp}),
the EW geometry is by no means essential to follow the calculations.
The reader who is merely interested in solutions to the dKP equation
(\ref{dkp}) could regard the constraint (\ref{constraint}),
the vanishing of differential forms 
(\ref{3_forms}), or the quadric ansatz (\ref{quadric}) as starting points.
\section{ The null case}
\label{snull}
The condition $\d u\wedge\ast\d u =0$ yields
\be
\label{constraint}
u_x(u_t-uu_x)=u_y^2.
\ee
We aim to find all solutions to the dKP equation (\ref{dkp}) subject
to the above constraint. 
\begin{prop}
The EW spaces {\em(\ref{EWdkp})} for which $|\d u|=0$ {\em(}or equivalently 
{\em(\ref{constraint})} holds {\em)} are locally given by one of the two forms
\begin{eqnarray}
\label{newEW}
h&=&\d y^2 +4C\d y\d t+4(yC'+t+2tCC')\d r\d t+C^2\d t^2,\nonumber\\
\om&=&\frac{4}{yC'+t+2tCC'}\d t,
\end{eqnarray}
or 
\begin{eqnarray}
\label{dKdV}
h&=&\d y^2+4(tP-1)\d t\d s,\nonumber\\
\om&=&4\frac{P}{tP-1}\d t,
\end{eqnarray}
where $C(r)$ and $P(s)$ are arbitrary functions, and $'=\d/\d r$.
The case {\em(\ref{dKdV})} 
is characterised by the existence of a symmetry $\p_y$.
\end{prop}
{\bf Proof.}
We first write  equations (\ref{constraint}), and (\ref{dkp})
as a system of first order
PDEs:
\begin{eqnarray*}
w_y&=&u_t-uu_x\\
w_x&=&u_y\\
0&=&u_xw_y-u_yw_x.
\end{eqnarray*}
Eliminating $w(x, y, t)$ from the first two equations yields the dKP, 
and the third equation is equivalent to (\ref{constraint}). The condition
$w_x=u_y$ implies the existence of $H(x, y, t)$ such that
$u=H_x, w=H_y$. The dKP equation becomes
\be
\label{Heqn}
H_{yy}-H_{xt} +H_xH_{xx}=0,
\ee
and (\ref{constraint}) is equivalent to the Monge equation
\be
\label{monge}
H_{xx}H_{yy}-H_{xy}^2=0.
\ee
To solve (\ref{monge}) rewrite it as
\begin{eqnarray*}
0&=&\d H_x\wedge\d H_y\wedge\d t\\
\d H&=&H_x\d x+H_y\d y+H_t\d t
\end{eqnarray*} 
and perform the Legendre transform
\be
\label{legendreT}
r=H_x,\qquad F(t, y, r)=H(t, y, x(t, y, r))-rx(t, y, r),
\ee
so that 
\[
H_y=F_y, \qquad H_t=F_t, \qquad x=-F_r.
\]
Differentiating these relation with respect to $r$ 
yields
\be
\label{identities}
H_{xx}=-\frac{1}{F_{rr}}, \qquad
H_{xy}=-\frac{F_{yr}}{F_{rr}}, \qquad
H_{yy}=F_{yy}-\frac{{F_{ry}}^2}{F_{rr}}.
\ee
Equation (\ref{monge}) becomes $F_{yy}=0$ with
the general solution
\be
\label{solution}
F=yA(r, t)+B(r, t).
\ee
Now we shall move on to equation (\ref{Heqn}).
We  write it as
\[
0=\d H_x\wedge\d x\wedge\d y-\d\Big(\frac{{H_x}^2}{2}\Big)\wedge\d t\wedge\d y+
\d H_y\wedge\d x\wedge \d t
\]
and perform the Legendre transform (\ref{legendreT}) 
to $(t, y, r)$ coordinates. We need to consider two cases:
\begin{enumerate}
\item $\d H_y\wedge\d x\wedge\d t\neq 0$ (or $u_y\neq 0$).
The resulting equation is
\[
F_{rt}+F_{yy}F_{rr}-F_{ry}^2=r.
\]
Substituting the solution (\ref{solution}) yields
\[
A=a(r)+c(t), \qquad B=\frac{r^2t}{2}+t\int (a')^2\d r+b(t),\qquad
\mbox{where}\;'=\frac{\d}{\d r}.
\]
To write down the EW structure (\ref{EWdkp}) define $C(r)=a'(r)$, and use
(\ref{identities}) together with $u=H_x$. This yields 
(\ref{newEW}).
\item $\d H_y\wedge\d x\wedge\d t= 0$ (or $u_y=0$).
The resulting equation $r=F_{rt}$
implies 
\[
F=\frac{r^2t}{2}+m(t)+n(r).
\]
Therefore
\[
h=\d y^2+4(t+n_{rr})\d r\d t, \qquad \om =\frac{4}{t+n_{rr}}.
\]
If we now define $s(r)$ by $\d s=(n_{rr})^{-1}\d r$, and 
put $P(s)=(n_{rr})^{-1}$ then the EW structure becomes
(\ref{dKdV}).
This is precisely the EW structure on the $S^{1}$ bundle over a
two-dimensional indefinite Weyl manifold constructed in \cite{DMT00}
out of solutions to $u_t-uu_x=0$.
\end{enumerate}
\koniec
\section{The generic case: generalised Nahm's equations}
\label{Nahm}
If $|\d u|\neq 0$ then (\ref{gdkp}) implies that there exist
holomorphic (or smooth if we seek real solutions) symplectic form 
$\om=f(p, q)\d p \wedge\d q$ such that
\[
\ast\d u=u_x\d y\wedge\d x+(u_t-2uu_x)\d t\wedge\d y+2u_y\d t\wedge\d x
=\om.
\]
This enables us to rewrite the dKP equation in terms of differential
forms. Define three three-forms in an open set of $\C^6$ (or $\R^6$) by
\begin{eqnarray}
\label{3_forms}
\Om_1&=&\d u\wedge\d y\wedge \d t+\om\wedge\d t\nonumber,\\
\Om_2&=&\d u\wedge\d x\wedge\d y-\d(u^2)\wedge\d y\wedge\d t+\d
x\wedge\om,\\
\Om_3&=&2\d u\wedge\d t\wedge\d x-\om\wedge\d y\nonumber.
\end{eqnarray}
The dKP equation with $|\d u|\neq 0$ is equivalent to 
\[
\Om_1=0,\qquad\Om_2=0,\qquad \Om_3=0.
\]
Selecting a three dimensional surface (an integral manifold) 
in $\C^6$ with $(x, y, t)$ as the
local coordinates, and eliminating $p, q$ by cross-differentiating
would lead back to (\ref{dkp}). We are however free to make another
choice, and use $(p, q, u)$ as the coordinates. This leads to the 
following set of first order PDEs:
\begin{eqnarray}
\label{sdiff1}
\dot{t}&=&\{t, y\}\nonumber\\
\dot{x}&=&\{y, x\}+2u\{y, t\}\\
\dot{y}&=&2\{t, x\}\nonumber,
\end{eqnarray} 
where the Poisson structure $\{\;,\;\}$ is induced by $\om$, and $
{\dot{\;}}=\p/\p u$. 

Alternatively defining $s=x+ut$ yields
\begin{eqnarray}
\label{sdiff3}
\dot{s}&=&\{y, s\}+t\nonumber\\
\dot{t}&=&\{t, y\}\\
\dot{y}&=&2\{t, s\}\nonumber.
\end{eqnarray}
\subsection{Generalised Nahm's equation}
Equations (\ref{sdiff3}) are contained in a generalisation of the Nahm
system proposed in \cite{C01}.
Consider an n-dimensional complex Lie group $G$ with a Lie algebra 
${\bf g}$. 
Let $[\;,\;]_{\bf g}$ be the Lie bracket in ${\bf g}$, and 
let $\Phi:\C\longrightarrow {\bf g}\otimes\C^3$, so that
$\Phi_i=\Phi_{i\alpha}(u)\sm^{\alpha}$ where
$u$ is a coordinate on $\C$,
$\sm^{\alpha}, \alpha=1, ..., n$ are generators of ${\bf g}$, and $i=1,2,3$.
The Nahm equations are
\be
\label{standart}
\dot{\Phi}_i=\frac{1}{2}\varepsilon_{ijk}[\Phi_j, \Phi_k]_{\bf g}.
\ee
(The summation convention is assumed, except where  stated.)
To generalise (\ref{standart})
introduce a
a $3\times 3$ trace-free matrix $B(u)$ such that
\be
\label{david}
\dot{\Phi}_{i}=\frac{1}{2}\varepsilon_{ijk}[\Phi_j, \Phi_k]_{\bf g}+B_{ij}\Phi_j.
\ee
Calderbank  \cite{C01} claims that equations (\ref{david})
are integrable iff $B(u)$ satisfies the matrix-Riccati equation
\[
\dot{B}=2(B^2)_0,
\]
where $({B^2})_0$ is the trace-free part of $B^2$.
Consider the  system (\ref{david}) with ${\bf g}={\bf sl}(2, \C)$, and rewrite
$\Phi_i$ as a symmetric matrix 
$\Phi_{AB}$ where $A, B$ are two-dimensional indices, and
$\Phi_{AB}$ is symmetric in $A$ and $B$
\[
\Phi_i\longrightarrow\Phi_{AB}(u)=\frac{1}{\sqrt{2}}\left(\begin{array}{cc}
\Phi_1+\Phi_2&\Phi_3\\
\Phi_3&\Phi_1-\Phi_2
\end{array}\right)=
\left(\begin{array}{cc}
2S(u)&-Y(u)\\
-Y(u)&2T(u)
\end{array}\right).
\]
Equations (\ref{david}) become
\be
\label{spinor_nahm}
\dot{\Phi}_{AB}=\frac{1}{2}\varepsilon^{CD}[\Phi_{AC}, \Phi_{BD}]
+\Phi_{CD} {\beta^C}_{A}{\beta^D}_B.
\ee
Here $\varepsilon^{CD}$ is a constant symplectic structure
on $\C^2$ with $\varepsilon^{01}=1$, and $\beta:\C\longrightarrow {\bf sl}(2,
\C)$.

To write down a Lax pair for (\ref{spinor_nahm}) we shall work with a
homogeneous space for $SL(2, \C)$. This is the space ${\cal F}$ of 
all pairs $(\pi^A, \eta^B)$, such that
$\varepsilon_{AB}\pi^{A}\eta^B=1$. 
We shall restrict ourself to (\ref{spinor_nahm}) 
for which $\beta$ is nilpotent (and therefore constant as a consequence of the
matrix Riccati equation),
or ${\beta^A}_B=-\iota^A\iota_B$, where $\iota_B=(-1, 0)$ and
$\iota^{B}=\varepsilon^{AB}\iota_B=(0, 1)$. 
Let
\begin{eqnarray*}
{\cal A}&=&\frac{\Phi_{AB}\pi^A\pi^B}{{(\pi_C\iota^C)}^4}=
2(S+\l Y+\l^2 T), \\
{\cal B}&=&-\Phi_{AB}\pi^A\eta^B=Y+2\l T\qquad\mbox{where}\qquad\eta^A=
\frac{\iota^A}{\pi_{B}\iota^B}.
\end{eqnarray*}
Here $\lambda=\pi_{0}/\pi_{1}$ is the affine coordinate on $\CP^1$,
and we put $\pi_{1}=1$.

The Lax pair for (\ref{spinor_nahm}) is in this case given by
\be
\label{lax_pair}
L_0=\frac{\p}{\p \l}-{\cal A}, \qquad
L_1=\frac{\p}{\p u}-{\cal B}.
\ee
The integrability $[L_0, L_1]=0$ of this distribution is equivalent to
\begin{eqnarray}
\label{sl2nahm}
\dot{S}&=&[Y, S]+T,\nonumber\\
\dot{T}&=&[T, Y],\\
\dot{Y}&=&2[T, S]\nonumber,
\end{eqnarray}
which is   (\ref{spinor_nahm}).\\
{\bf Remark.} The anonymous referee has pointed out that the
generalised Nahm's equations (\ref{david}) arise as symmetry
reductions of anti-self-dual Yang-Mills equations on $\C^4$ 
by three-dimensional abelian sub-groups of $PGL(4, \C)$.
When the gauge group is $SL(2, \C)$, the corresponding reductions are
the six Painlev\'e equations \cite{MW96}. 
\subsection{sdiff$(\Sm)$ generalised Nahms equation and dKP}
Now assume that ${\bf g}$ is the infinite-dimensional Lie algebra 
sdiff$(\Sm)$ of holomorphic symplectomorphisms of a two-dimensional complex 
symplectic manifold $\Sm$ with local holomorphic 
coordinates $p, q$ and the holomorphic symplectic
structure $\om$. Elements of
sdiff$(\Sm)$ are represented by the Hamiltonian vector fields $X_H$
such that
\[
X_H\hook \om=\d H
\] 
where $H$ is a $\C$-valued function on $\Sm$.
The Poisson algebra of functions which we are going to use 
is homomorphic to sdiff$(\Sm)$. 
We shall make the following replacement 
\[
[\;,\;]_{\bf g}\longrightarrow \{\;,\;\}. 
\]
in formulae (\ref{spinor_nahm},\ref{sl2nahm}).
The components of
$\Phi_{AB}$ are therefore regarded as Hamiltonians generating the 
symplectomorphisms of $\Sm$, and the generalised Nahm equations 
(\ref{spinor_nahm})
with 
\be
\label{solutions}
\Phi_{AB}(u, p, q)=\left(\begin{array}{cc}
2s(u, p, q)&-y(u, p, q)\\
-y(u, p, q)&2t(u, p, q)
\end{array}\right),\qquad {\beta^A}_B= \left(\begin{array}{cc}
0&1\\
0&0\end{array}\right).
\ee
yield (\ref{sdiff3}).

The Lax pair for (\ref{sdiff3}) is therefore given by
\be
\label{la}
L_0=\frac{\p}{\p \l}+{X_{H_{\cal A}}}, \qquad L_1=\frac{\p}{\p u}+
X_{H_{\cal B}},
\ee
where $X_{H_{\cal A}}$, and  $X_{H_{\cal B}}$ are Hamiltonian vector fields with 
Hamiltonians
\[
H_{\cal A}=2(s+\l y+ \l^2 t), \qquad H_{\cal B}=y+2\l t.
\]
To verify this claim we notice that
the condition 
$[L_0, L_1]=0$ and the standard relation
$[X_f, X_g]=-X_{\{f, g\}}$ yield
\[
\p_u{H_{\cal A}}-\p_\l H_{\cal B}+\{H_{\cal A}, H_{\cal B}\}=0.
\]
Therefore
\[
\dot{s}+\l\dot y+\l^2\dot{t}=t-\{s, y\}-2\l\{s, t\}-\l^2\{y, t\}
\]
which is (\ref{sdiff3}).
\subsection{${\bf sl}(2, \C)$ generalised Nahm's equation  and 
the central cone ansatz.}
\label{sCONE}
In this section we shall establish the following result
\begin{prop}
\label{CCONE}
The solutions of the 
$SL(2, \C)$ generalised
Nahm's equations {\em(\ref{sl2nahm})}  
correspond to solutions of the dKP equation 
{\em(\ref{dkp})} 
which are constant on a central quadric, in the sense that
$u(x, y, t)$ is determined implicitly by an equation of the form
\be
\label{cone}
Q(x, y, t, u)={X}^T{\bf M}(u){X}=C,
\ee
where $C$ is a constant, $X^T=(x, y, t)$, and ${\bf M}(u)$ is a symmetric matrix whose
components depend on $u$.
\end{prop}
{\bf Proof.}
Let 
\[
\sigma=\left(\begin{array}{cc}
1/2&0\\
0&-1/2
\end{array}
\right ),\qquad
\sigma_+=\left(\begin{array}{cc}
0&1\\
0&0
\end{array}
\right ),\qquad
\sigma_-=\left(\begin{array}{cc}
0&0\\
1&0
\end{array}
\right )
\]
be a basis of ${\bf sl}(2, \C)$ such that
\[
[\sigma_+, \sigma_-]=2\sigma,\qquad
[\sigma, \sigma_\pm]=\pm\sigma_\pm.
\]
Put 
\begin{eqnarray}
\label{system}
S&=&a_{11}\sigma_+ +a_{12}\sigma +a_{13}\sigma_-\nonumber\\
Y&=&a_{21}\sigma_+ +a_{22}\sigma +a_{23}\sigma_-\\
T&=&a_{31}\sigma_+ +a_{32}\sigma +a_{33}\sigma_-\nonumber,
\end{eqnarray}
where the coefficients $a_{ij}$ of 
the $3\times 3$ matrix ${\bf a}={\bf a}(u)$
are functions of $u$ (in the next section they will be expressed by 
solutions to Painlev\'e I, or Painlev\'e II).
We can construct an explicit embedding of 
${\bf sl}(2, \C)\subset$sdiff$(\Sm)$ where by considering  
the infinitesimal linear or M\"obius action of $SL(2, \C)$ on
$\Sm$. The Hamiltonians corresponding to matrices in ${\bf sl}(2, \C)$
are defined up to the addition of a function of $u$.
Let 
\[
\left(\begin{array}{cc}
A&B\\
C&D
\end{array}
\right)\in SL(2, \C).
\]
\begin{itemize}
\item $\Sm=\C^2$.
The linear action 
\[
(p, q)\longrightarrow(Ap+Bq, Cp+Dq)
\]
preserves $\om =\d p\wedge\d q$ and is generated by
\be
\label{repl1}
H_\sm=\frac{pq}{2}, \qquad
H_{\sm_+}=-\frac{q^2}{2},\qquad H_{\sm_-}=\frac{p^2}{2}.
\ee
The Hamiltonians satisfy the algebraic constraint
\be
\label{aconstraint1}
{(H_\sm)}^2+H_{\sm_+}H_{\sm_-}=0.
\ee
\item $\Sm=\CP^1\times\CP^1$. The group $SL(2, \C)$ acts on one
Riemann sphere by M\"obius transformation, and on the other by the
inverse
\[
(p, q)\longrightarrow\Big(\frac{Ap+B}{Cp+D}, \frac{Dq-C}{-Bq+A}\Big).
\]
Here $p$ and $q$ are independent holomorphic coordinates on both
Riemann spheres, and the action preserves $\om=(1+pq)^{-2}\d p\wedge\d
q$. The corresponding Hamiltonians
\be
\label{repl2}
H_{\sm}=\frac{1}{2}\frac{pq-1}{pq+1},\qquad
H_{\sm_+}=\frac{q}{1+pq},\qquad H_{\sm_-}=\frac{p}{1+pq}
\ee
satisfy
\be
\label{aconstraint2}
{(H_\sm)}^2+H_{\sm_+}H_{\sm_-}=\frac{1}{4}.
\ee
\end{itemize}
In both case the matrices $(T, Y, S)$ are replaced by functions
\[
t(u, p, q),\qquad y(u, p, q), \qquad s(u, p, q)=x(u, p, q)+ut(u, p ,q).
\]
Regard (\ref{system}) as a system of equations for $(H_{\sm_+}, H_{\sm},
H_{\sm_-})$, and assume for the moment that $\det{({\bf a}(u))}\neq 0$.
The algebraic constraints (\ref{aconstraint1},\ref{aconstraint2})
imply
\begin{eqnarray}
\label{detconstr}
&&\det{\left(\begin{array}{ccc}
a_{11}&x+ut&a_{13}\\
a_{21}&y&a_{23}\\
a_{31}&t&a_{33}
\end{array}
\right )}^2+\det{\left(\begin{array}{ccc}
x+ut&a_{12}&a_{13}\\
y&a_{22}&a_{23}\\
t&a_{32}&a_{33}
\end{array}
\right )}
\det{\left(\begin{array}{ccc}
a_{11}&a_{12}&x+ut\\
a_{21}&a_{22}&y\\
a_{31}&a_{32}&t
\end{array}
\right )}\nonumber\\
&&=C\;\det{({\bf a}(u))}^2,
\end{eqnarray}
(where $C=\mbox{const}$) which is yields to  (\ref{cone}).
\koniec
{\bf Remarks.} 
\begin{itemize}
\item 
Note that a similarity transformation
\[
T\rightarrow g^{-1}Tg,\qquad
Y\rightarrow g^{-1}Yg,\qquad
S\rightarrow g^{-1}Sg,
\]
where $g=g(u)\in SL(2, \C)$ does not change (\ref{detconstr}) or (\ref{cone}). 
This can be checked by an explicit (MAPLE) calculation. It also follows
from the fact that the transformation $H_{K}\rightarrow H_{\hat{K}}$
defined by the following diagram
\begin{eqnarray*}
K 
&\longrightarrow &H_{K}\\
\Big\downarrow & &\Big\downarrow\\
g^{-1}Kg &\longrightarrow& H_{\hat{K}},
\end{eqnarray*}
where
\begin{eqnarray*}
K&=&a\sm_+ +b\sm+c\sm_-\in \mbox{Map}(\C, {\bf sl}(2, \C)),\\
H_{K}&=&aH_{\sm_+} +bH_{\sm}+cH_{\sm_-}\in\mbox{Map}(\C\times\Sm,
\C)
\end{eqnarray*}
is canonical and preserves the constraints 
(\ref{aconstraint1}, \ref{aconstraint2}).
\item 
Solutions to the dKP equations obtained from the ansatz
(\ref{cone}) with const$=0$ (a central cone) 
are invariant under scaling transformations generated by
$x\p_x+y\p_y+t\p_t$. In fact they form a subclass of all solutions
with this symmetry. This subclass is charactrerised by a 
`general quadric' ansatz
\[
{\cal Q}(\xi_1, \xi_2, u)=
\left(\begin{array}{cc}
\xi_{1}\\
\xi_{2}
\end{array}
\right )
\left(\begin{array}{cc}
m_{1}&m_{2}\\
m_{3}&m_{4}
\end{array}
\right )
\left(\begin{array}{cc}
\xi_{1}&\xi_{2}
\end{array}
\right )
+2m_5\xi_1+2m_6\xi_2+m_7=0,
\]
where $\xi_1=t/y, \xi_2=x/y$, and $m_{1}, ..., m_7$ depend on $u$.
\end{itemize}
\section{dKP from the isomonodromy problem associated to PI and PII}
\label{Spainleve}
In this section we shall relate the generalised Nahm equations 
(\ref{sl2nahm}) to the first two  Painlev\'e transcendents, and
construct the corresponding solutions to the dKP equation.

Consider the ODE
\be
\label{ODE}
\frac{\d\Psi}{\d \l}=\hat{\cal A}(\l)\Psi,\qquad\mbox{where}\qquad
\hat{\cal A}(\l)=2(\l^2{\hat{T}}+\l \hat{Y}+\hat{S}),
\ee
anc  $\hat{T}, \hat{Y}, \hat{S}$ are constant  elements of 
${\bf sl}(2, \C)$. Here  $\Psi(\l)$ is a matrix fundamental solution (it
takes values in $SL(2, \C)$). 
Consider a one-parameter   deformation 
$(\hat{T}(z), \hat{Y}(z), \hat{S}(z))$ of this ODE.
It is known \cite{JM81} that monodromy data around the 
fourth-order pole at $\l=\infty$ remains constant if
$\Psi(\l, z)$ satisfies
\be
\label{monodromlax}
\frac{\p\Psi}{\p \l}=\hat{\cal A}(\l)\Psi,\qquad
\frac{\p \Psi}{\p z}=\hat{\cal B}(\l)\Psi, 
\ee
for a certain matrix ${\hat{\cal B}}$ whose exact form depends on
whether $\hat{T}$ is diagonalizable or nilpotent. 
For nilpotent (diagonalizable) ${\hat T}$
the compatibility conditions $\Psi_{\l z}=\Psi_{z \l}$, or
\[
[\p_\l-\hat{\cal A}(\l), \p_z-\hat{\cal B}(\l)]=0
\]
reduce to the first (the second) Painlev\'e equation. 
\begin{itemize}
\item {\bf Diagonalizable} $\hat{T}$.
The explicit parametrisation of  $\hat{T}(z), \hat{Y}(z), \hat{S}(z)$ 
can be chosen so that
\[
\hat{T}=\sm,\qquad\hat{Y}=\frac{\theta}{2}\sm_+-\frac{v}{\theta}\sm_-, 
\qquad\hat{S}=\Big(v+\frac{z}{2}\Big)\sm-\frac{\theta w}{2}\sm_+
-\frac{wv-\alpha+1/2}{\theta}\sm_-.
\]
where $w, v$,  and $\theta$ are functions of $z$, and $\alpha$ is a
parameter. 
It this case $\hat{\cal B}(\l)=\l\hat{T}+\hat{Y}$, and 
the compatibility conditions for (\ref{monodromlax})
are
\[
\frac{\d w}{\d z}=v+w^2+\frac{z}{2},\qquad
\frac{\d v}{\d z}=-2wv-\frac{1}{2}+\alpha, \qquad
\frac{\d}{\d z
}\ln{\theta}=-w,
\]
and finally
\be
\label{painleve2}
\frac{\d^2 w}{\d z^2}=2w^3+zw+\alpha,
\ee
which is Painlev\'e II \cite{Ince}.
\item {\bf Nilpotent} $\hat{T}$. Now
\[
\hat{T}=\frac{1}{2}\sm_+,\qquad\hat{Y}=\frac{w}{2}\sm_+ +2\sm_-,
\qquad \hat{S}=\frac{2w^2+z}{4}\sm_+ -v\sm -2w\sm_-,
\]
where $w$ and $v$ depend on $z$. The matrix ${\hat{\cal B}}$
is given by 
$
\hat{\cal B}(\l)=\l\hat{T}+(1/2)\hat{Y}+ 
(w/2)\sm_+$, and the compatibility conditions for (\ref{monodromlax})
are 
\[
\frac{\d w}{\d z}=v,\qquad\frac{\d v}{\d z}=6w^2+z,
\]
and
\be
\label{painleve1}
\frac{\d^2 w}{\d z^2}=6w^2+z,
\ee
which is Painlev\'e I \cite{Ince}.
\end{itemize}
Replace $z$ by $2u$, and put $\hat{\cal B}=\hat{\cal B}_1/2$.
The matrices $\hat{\cal A}, \hat{\cal B}_1$ are defined up to gauge transformations
\[
\hat{\cal A}\longrightarrow {\cal A}=g^{-1}\hat{\cal A}g-g^{-1}\p_\l g,
\qquad 
\hat{\cal B}_1\longrightarrow {\cal B}=g^{-1}\hat{\cal B}_1g-g^{-1}\p_u g,
\]
where $g=g(u, \l)\in SL(2,\C)$.
\begin{itemize}
\item {\bf Diagonalizable}  $\hat{T}$.
Choose $g=g(u)$ such that $g^{-1}\p _u g=g^{-1} \hat{Y} g$, (which
implies $g=\exp{(\int\hat{Y}\d u)}$).
\item {\bf Nilpotent} $\hat{T}$.
Choose $g=g(u)$ such that $g^{-1}\p _u g=w\sm_+$, (so
$g=\exp{(-f\sm_+)}$, where $\d f/\d u=-w$).
\end{itemize}
After all these transformations
\[
\hat{\cal A}\longrightarrow
{\cal A}=
2(\l^2{T}+\l Y+ S),\qquad \hat{\cal B}\longrightarrow {\cal B}= 2\l T+Y,
\]
where $T=g^{-1}\hat{T}g, Y=g^{-1}\hat{Y}g, S=g^{-1}\hat{S}g$
and the isomonodromy Lax pair  (\ref{monodromlax}) 
becomes the Lax pair
(\ref{lax_pair}) for the generalised Nahm equations 
(\ref{spinor_nahm}) with nilpotent $\beta$ (i.e. (\ref{sl2nahm})).
Moreover making the replacements (\ref{repl1},\ref{repl2}),  and using the results
of the last subsection we deduce that any solution to Painlev\'e I, or
II
yields a solution to the dKP constant on a central quadric.

Keeping in mind the 1st remark after Proposition \ref{CCONE}
we shall  work out the details of this construction and find the 
matrix ${\bf M}(u)$ in (\ref{cone}).
\begin{itemize}
\item
For PI we verify that
$\det{({\bf a})}=v/2\neq 0$, and so the formula 
(\ref{detconstr}) can be used to read off the matrix ${\bf M}(u)$
\be
\label{cone1}
x^2+w^2y^2
-w\Big(\frac{{\dot w}^2}{4}-4w^3\Big)t^2-4xtw^2
+2wxy+\Big(\frac{{\dot w}^2}{4}-4w^3\Big)yt=\hat{C}{\dot w}^2,
\ee
where $w=w(u)$ satisfies the rescaled PI equation
\be
\label{rescaledPI}
\ddot{w}/4=6w^2+2u,
\ee
and the constant ${\hat C}$ can always be set to 0 or 1.
\item
For PII we find that $\det{({\bf a})}=wv-(\a-1/2)/2$.
Imposing the  constraint $\det{({\bf a})}=0$ 
on PII  leads to $v=0$ and $\a=1/2$.
We shall examine this case later, and now we shall concentrate on
the generic case $v\neq 0$. The quadric becomes
\begin{eqnarray}
\label{cone2}
&&{x}^{2}v-
{y}^{2}w\left (wv-(\a-1/2)\right )+\frac{1}{2}\,{t}^{2}\left (\left (\a-1/2
\right )^{2}+4\,wv\left (wv-(\a-1/2)\right )+2\,{v}^{3}\right )
\nonumber\\
&&+xy\left (\a-1/2\right )-ytv\left (\a-1/2\right )-2\,tx{v}^{2}=
\hat{C}(2wv-(\alpha-1/2))^2,
\end{eqnarray}
where
\be
\label{PIIv}
v=\frac{1}{2}\dot{w}(u)-w(u)^{2}-u,
\ee
\be
\label{rescalPII}
\frac{1}{4}\ddot{w}=2w^3+2wu+\a,
\ee
and ${\hat C}=0$, or $1$.

We still have to consider $\det{(\bf{a})}=0$ (or $v=0, \a=1/2$).
Putting 
$
w(u)=-{\dot{f}}/{(2f)}
$
reduces (\ref{rescalPII}) to the Airy equation 
\[
\ddot{f}+4uf=0.
\]
We shall recover $u(x, y, t)$ explicitly from the matrices
$(\hat{T}, \hat{Y}, \hat{S})$.
Applying the formula
\[
e^gKe^{-g}=K+[g, K]+\frac{1}{2!}[g,[g, K]]+...
\]
which holds for any square matrices $g$ and $K$
we find
\[
T=\sm+\frac{1}{2}[\int f\d u]\sm_+,\qquad
Y=\frac{1}{2}f\sm_+,\qquad
S=u\sm+\Big(\frac{1}{4}\dot{f}+\frac{u}{2}[\int f\d u]\Big)\sm_+,
\]
and the equations (\ref{sdiff1}) are satisfied with
\[
x=-\frac{\dot{f}q^2}{8},\qquad y=-\frac{fq^2}{4},\qquad t=\frac{pq}{2}-
\frac{[\int f\d u]q^2}{4}
\]
Differentiating the constraint $2xf=y\dot{f}$ we find that 
$u_t=xu_x+yu_y=0$, and $u(\xi)$ (where $\xi=x/y$) satisfies
\[
2\xi^2u_{\xi}+2uu_{\xi}+1=0.
\]
This ODE is reducible to a special kind of Abel's equation.
\end{itemize}
We can summarise the results of this section in the following
theorem
 \begin{theo}
\label{mainth}
The $SL(2, \C)$ generalised Nahm equations {\em(\ref{sl2nahm})}
are equivalent to Painlev\'e I if $T$ is nilpotent and to Painlev\'e II if $T$ is
diagonal. The corresponding solutions to the dKP equations are constant
on a central quadrics 
$Q(x, y, t, u)={X}^T{\bf M}(u){X}=C$ where $X^T=(x, y, t)$, and ${\bf M}(u)$ 
can be read-off form {\em(\ref{cone1})} and {\em(\ref{cone2})}.
\end{theo}
\section{The quadric ansatz}
\label{quadricAN}
The solutions to dKP found in Proposition \ref{CCONE} have a form which is 
reminiscent of some solutions of the $SU(\infty)$ Toda field equations
constant on central quadrics found in \cite{T95}.

We shall now  present a converse to Proposition \ref{CCONE}, 
and show that the existence of Painlev\'e solutions to dKP 
follows from what we may call the quadric ansatz, which may have
a wider utility.
This ansatz can be made whenever we have a non-linear PDE of the form
\be
\label{PDE}
\frac{\p}{\p x^j}\Big( b^{ij}(u)\frac{\p u}{\p x^i}\Big)=0,
\ee
where $u$ is a function of coordinates $x^i, i=1, ..., n$.
The ansatz is to seek solutions constant on central quadrics 
or equivalently to seek a matrix ${\bf M}(u)=(M_{ij}(u))$ so that
a solution of equation (\ref{PDE})  is determined implicitly as in 
(\ref{quadric}) by
\be
\label{q2}
Q(x^i, u)=M_{ij}(u)x^ix^j=C.
\ee
(This ansatz is motivated by the work of Darboux \cite{Darboux} orthogonal 
curvilinear coordinates.)
We differentiate (\ref{q2}) implicitly to find
\be
\label{ux}
\frac{\p u}{\p x^i}=-\frac{2}{\dot{Q}}M_{ij}x^j,\qquad
\mbox{where}\qquad \dot{Q}=\frac{\p Q}{\p u}.
\ee
Now we substitute this into (\ref{PDE}) and integrate once with
respect to $u$. Introducing $g(u)$ by
\be
\label{gu}
\dot{g}=\frac{1}{2}b^{ij}M_{ij}=\frac{1}{2}\mbox{trace}\;{(\bf bM)}
\ee
we obtain
\[
(g\dot{M}_{ij}-M_{ik}b^{km}M_{mj})x^ix^j=0,
\]
so that as a matrix ODE
\be
\label{mode}
g\dot{\bf{M}}={\bf{MbM}}.
\ee
This equation simplifies if written in terms of another matrix
${\bf N}(u)$ where
\be
\label{NM}
{\bf N}=-{\bf M}^{-1}
\ee
for then
\be
\label{gnb}
g\dot{\bf{N}}={\bf b},
\ee
and $g$ can be given in terms of $\triangle=\det{({\bf{N}})}$ by
\be
\label{const}
g^2\triangle=\zeta=\mbox{constant}.
\ee
Restricting to three dimensions with
$(x^1, x^2, x^3)=(x, y, t)$, 
the $SU(\infty)$ Toda field equation
\[
u_{xx}+u_{yy}+{(e^u)}_{tt}=0
\]
is given by (\ref{PDE}) with 
\[
{\bf b}(u)=
\left (
\begin{array}{ccc}
1&0&0\\
0&1&0\\
0&0&e^u
\end{array}
\right )
\]
and as was shown in \cite{T95}, in this case (\ref{gnb})
can be reduced to Painlev\'e III.
\begin{theo}
\label{Qquadric}
Solutions to the dKP equation {\em(\ref{dkp})} constant on the central
quadric {\em(\ref{quadric})} are implicitly given by
solutions to Painlev\'e I or II:
\begin{itemize}
\item
If $({\bf M}^{-1})_{33}\neq 0$ then
\begin{eqnarray}
\label{PIIquadric}
&&{x}^{2}v-
{y}^{2}w\left (wv-(\a-1/2)\right )+\frac{1}{2}\,{t}^{2}\left (\left (\a-1/2
\right )^{2}+4\,wv\left (wv-(\a-1/2)\right )+2\,{v}^{3}\right )\nonumber\\
&&+xy\left (\a-1/2\right )-ytv\left (\a-1/2\right )-2\,tx{v}^{2}
=\hat{C}(2wv-(\alpha-1/2))^2,
\end{eqnarray}
where $\a$ is a constant, $v$ is given by 
{\em(\ref{PIIv})}, and  $w$ is a solution to the rescaled Painlev\'e II   
{\em(\ref{rescalPII})}.
\item 
If $({\bf M}^{-1})_{33}=0$ and $({\bf M}^{-1})_{23}\neq 0$  then
\be
\label{PIquadric}
x^2+w^2y^2
-w\Big(\frac{{\dot w}^2}{4}-4w^3\Big)t^2-4xtw^2
+2wxy+\Big(\frac{{\dot w}^2}{4}-4w^3\Big)yt=\hat{C}\dot{w}^2,
\ee
where $w(u)$ satisfies the rescaled Painlev\'e I
{\em(\ref{rescaledPI})}.
\item 
If $({\bf M}^{-1})_{33}=({\bf M}^{-1})_{23}= 0$  then
\be
\label{degenerate}
\frac{{y}^{2}}{4}+\left(\sin{(u)}^{3}\cos(u)-u\sin(u)^{2}
+{\gamma}^{2}\cos(u)^{4}\right){t}^{2}
-\sin(u)^{2}tx-\gamma\cos(u)^{2}ty=
\hat{C} \tan(u)^{2}
\ee
where $\gamma$ is a constant.
\end{itemize}
The constant $\hat{C}$ can always be set to $0$ or $1$. 
\end{theo}
{\bf Proof.}
For the dKP equation we have (\ref{PDE}) with
\be
\label{bdKP}
{\bf b}(u)=
\left (
\begin{array}{ccc}
-u&0&1/2\\
0&-1&0\\
1/2&0&0
\end{array}
\right ).
\ee
(Note that $-{\bf b}(u)$ is the inverse of $h$ in (\ref{EWdkp})).
Equation (\ref{gnb}) implies that ${\bf N}(u)$ can
be written as
\be
\label{NdKP}
{\bf N}(u)=\left (
\begin{array}{ccc}
Y&\beta&Z\\
\beta&X&\epsilon\\
Z&\epsilon&\phi
\end{array}
\right )
\ee
where $\beta, \epsilon$ and $\phi$ are constants, while
\be
\label{XYZ}
\dot{Y}=-g^{-1}u,\qquad \dot{X}=-g^{-1},\qquad
\dot{Z}=\frac{1}{2}g^{-1}.
\ee
Now
\be
\label{eta}
\eta=X+2Z=\mbox{constant}
\ee
and we can eliminate $Z$ in favour of $X$ and another constant.
To eliminate $Y$ in favour of $X, \dot{X}$ and another constant
we use (\ref{const}) and (\ref{eta}):
\be
\label{triangle}
\triangle=(\phi X-\epsilon^2)Y-\frac{1}{4}X^3+\frac{1}{2}\eta X^2
-X(\frac{1}{4}\eta^2+\epsilon\beta)-\beta^2\phi+\epsilon\beta\eta
=\zeta \dot{X}^2
\ee
so that, provided $\phi$ and $\epsilon$ are not both zero
\be
\label{eqY}
Y=(\phi X-\epsilon^2)^{-1}(\zeta \dot{X}^2
+\frac{1}{4}X^3 -\frac{1}{2}\eta X^2
+X(\frac{1}{4}\eta^2+\epsilon\beta)+\beta^2\phi-\epsilon\beta\eta).
\ee
If $\phi=\epsilon=0$ then (\ref{triangle}) is a first order
ODE for $X$
\[
\zeta\dot{X}^2=-\frac{1}{4}X(\eta-X)^2
\]
which is readily solved yielding (\ref{degenerate}).
Otherwise we obtain $\dot{Y}$ from (\ref{eqY}) and substitute it into 
the equation 
\be
\label{yux}
\dot{Y}-u\dot{X}=0
\ee
obtained form (\ref{XYZ}).
This is the desired second order ODE for $X$ which we shall
investigate below.

First we use some freedom of redefinition to absorb some of the
constants. Note that the linear change of coordinates
\be
\label{redefinition}
\left (
\begin{array}{ccc}
\hat{x}\\
\hat{y}\\
\hat{t}
\end{array}
\right )=
\left (
\begin{array}{ccc}
1&c_1&c_2\\
0&1&2c_1\\
0&0&1
\end{array}
\right )
\left (
\begin{array}{ccc}
{x}\\
{y}\\
{t}
\end{array}
\right ),\qquad \hat{u}=u-c_1^2-\frac{1}{2}c_1-\frac{1}{2}c_2
\ee
is a symmetry of the dKP equation \cite{DMT00}.
This transformation preserves the quadric ansatz but changes
the constants as follows
\begin{eqnarray}
\label{TRasnf}
\hat{\phi}&=&\phi\nonumber\\
\hat{\epsilon}&=&\epsilon+2c_1\phi\\
\hat{\beta}&=&\beta+(c_2+2c_1^2)\epsilon+2c_1c_2\phi+c_1\eta\nonumber\\
\hat{\eta}&=&\eta+6c_1\epsilon+(4c_1^2+2c_2)\phi\nonumber
\end{eqnarray}
\begin{itemize}
\item
If $\phi\neq 0$, we can use this transformation to set
$\epsilon=\eta=0$.
Now introduce functions $v(u)$ and $w(u)$ and a constant $\alpha$ by
\be
\label{XYa}
X=-2\phi v,\qquad Y=2\phi w(vw -(\alpha-\frac{1}{2}))+\phi v^2,\qquad
\alpha=\frac{1}{2}+\frac{\beta}{\phi}
\ee
to find that the equations (\ref{triangle})  and (\ref{yux}) with 
$\zeta=-\phi/16$ are equivalent to (\ref{rescalPII}) with $v$ given by 
(\ref{PIIv}). This case reduces to Painlev\'e II.
The quadric (\ref{quadric}) can be written as (\ref{PIIquadric}),
which is identical to the previously obtained  
(\ref{cone2}).
\item
If $\phi=0$, we can use the transformation (\ref{TRasnf}) to set
$\beta=\eta=0$. Now introduce $w$ by
\be
\label{P1w}
X=2\epsilon w,\qquad
Y=\frac{\epsilon}{2}\Big(\frac{\dot{w}^2}{4}-4w^3\Big)
\ee
to find that (\ref{triangle}) and (\ref{yux}) with $\zeta=-\epsilon/32$ are
equivalent to the rescaled Painlev\'e I (\ref{rescaledPI}).
The quadric (\ref{quadric}) is this time (\ref{PIquadric})
(which is  (\ref{cone1}).
\end{itemize}
If $\hat{C}\neq 0$, then we can set $\hat{C}=1$ by (possibly complex)
rescaling of $(x, y, t)$.
\koniec
\section{Examples}
\label{Sexamples}
All solutions to PI are transcendental \cite{Ni88}.
On the other hand for certain 
values of $\a$ PII admits particular solutions expressible
in terms of `known' functions.
For $\a=n\in \Z$ (\ref{rescalPII}) possesses rational
solutions, and for $\a=n+1/2$ there exists a class of 
solutions expressible by Airy
functions. 
 For example if $\a=1$, then (\ref{rescalPII}) is satisfied
by $w=-1/(2u)$. Now $v=-u$, and (\ref{cone2}) becomes cubic in u
\[
8t^2u^3+16xtu^2+(8x^2-4yt)u-(t^2+4xy)=c,\qquad c=\mbox{const}.
\]
The three roots of this cubic give three solutions to dKP. A root
which yields a real solution is
\begin{eqnarray}
u(x,y,t)&=&\frac{\sqrt[3]{A+12t\sqrt{B}}}{12t}+
\frac{6yt+4x^2}{3t\sqrt[3]{A+12t\sqrt{B}}}-\frac{2x}{3t},\\
A&=&144\,xyt+64\,{x}^{3}+108\,{t}^{3}-108\,tc,\nonumber\\
B&=&-96\,{y}^{3}t-48\,{y}^{2}{x}^{2}+216\,xy{t}^{2}+96\,{x}^{3}t
+81\,t^4\nonumber\\
&&-96\;\frac{cx^3}{t}-216\;cxy-162\;ct^2+81\;c^2.
\nonumber
\end{eqnarray}
Other rational solutions of PII (or rational matrices ${\bf M}$)
corresponding to different values of $\a$ can be generated by
the following B\"acklund transformation (see for example \cite{G00})
\[
w_{\alpha}
=-w_{(\alpha-1)}-\frac{2\alpha+1}{\dot{w}_{(\a-1)}+2w_{(\a-1)}^2+2u}.
\]
Here $w_{\a}=w_{\a}(u)$ is an arbitrary solution of PII with a
parameter $\a$. We can therefore formulate the following
\begin{col}
If the quadric {\em(\ref{quadric})} is rational in $u$ then it is 
necessarily of the
form {\em(\ref{PIIquadric})} and it can be found explicitly.
\end{col}
For example if $\a=2$ and $\hat{C}=0$ then  $w=1/(2u)-3u^2/(1+2u^3), 
v=-(1+2u^3)/(2u^2)$, and (\ref{cone2}) is
\begin{eqnarray*}
&&(8\,xt-2\,{y}^{2})+8\,{t}^{2}u-\left (6\,yt-4\,{x}^{2}\right ){u}^{2}-
\left (-4\,{y}^{2}-16\,xt\right ){u}^{3}-\left (12\,yx-11\,{t}^{2}
\right ){u}^{4}\\
&&-\left (-8\,{x}^{2}+12\,yt\right ){u}^{5}+16\,xt{u}^{6}
+8\,{t}^{2}{u}^{7}=0.
\end{eqnarray*}
\section{Final comments on associated geometries}
\label{Sgeometries}
We can now look back at our initial motivation and use the solutions
to the dKP equation we have obtained to construct the Einstein--Weyl 
structures. To write down the EW structures in a way which explicitly
depends on  solutions to PI or PII we shall change coordinates 
in (\ref{EWdkp}) to $(t, y, u)$, and regard $x$ as a function.
Both cones (\ref{PIquadric}) and (\ref{PIIquadric}) can be written as
\be
\label{quadratic}
x^2+2ax+b=0,
\ee
where $a=a(t, y, u), b=b(t, y, u)$. 
For example
\[
a=wy-2tw^2,\qquad b=w^2y^2+(\dot{w}^2/4-4w^3)(yt-wt^2)-\hat{C}\dot{w}^2.
\]
in the case of PI. Now we 
choose one root of (\ref{quadratic}), and differentiate
(\ref{quadratic}) with respect to $x$ to find an expression for $u_x$.
The resulting EW structure is
\be
\label{explEW}
h=\d y^2-4(u\d t-\d (a+\sqrt{a^2-b^2}))\d t,\qquad
\om= \frac{8\sqrt{a^2-b^2}}{2(a+\sqrt{a^2-b^2})a_u-b_u}\d t.
\ee

There is a class of four-dimensional anti-self-dual 
geometries associated to (\ref{explEW}).  
\begin{defi}
A null-K\"ahler structure on a four-manifold ${\cal M}$ 
consists of an inner
product $g$ of signature $(++--)$ and a real rank-two endomorphism $
N:T{\cal M}\rightarrow T{\cal M}$ parallel 
with respect to this inner product such that
\[
N^2=0,\qquad\mbox{and}\qquad g(NX, Y)+g(X, NY)=0
\]
for all $X, Y \in T{\cal M}$.
A null-K{\"a}hler structure is anti-self-dual (ASD)
if the self-dual part of the Weyl tensor vanishes.
\end{defi}
In \cite{D02} it has  been shown that all ASD  null K\"ahler structures
with a symmetry which preserves $N$ are locally given by
solutions to the dKP equation (\ref{dkp})
and its linearisation.
\be
\label{lindKP}
V_{yy}-V_{xt}+(uV)_{xx}=0.
\ee
Given $u$ and $V$ the associated ASD null K\"ahler metric with a 
symmetry $\p_z$ is given by
\be
\label{Vag}
g=V(\d y^2-4\d x\d t-4u\d t^2)- V^{-1}(\d z+\beta)^2,
\ee
where the one-form $\beta$ is a solution to the monopole equation
\[
\ast(\d V+(1/2)\om V) =\d\beta,
\]
and the null-K\"ahler two-form is $\Om:=g(N,\; )=\d z\wedge\d t$.

 Now assume that  $u(x, y, t)$ is given in terms of PI or PII, so that
$(h, \om)$ are of the form (\ref{explEW}). A general $V$ will then lead to 
metrics (\ref{Vag}) with just one symmetry. If $V=\mbox{const}.u_x$ then
(\ref{lindKP}) reduces to (\ref{dkp}) and $g$ is (pseudo) hyper-K\"ahler.
Another particular solution of the monopole 
equation is picked out by the quadric ansatz:
If u is a solution of dKP  given by the
quadric ansatz (\ref{quadric}) then $V=\p u/\p C$ satisfies
(\ref{lindKP}). However by implicit differentiation of 
(\ref{quadric}) this is just
\[
V=\frac{1}{X^{T}\dot{\bf M}X}.
\]

 There is  a class of solutions to (\ref{lindKP}) such that 
the resulting ASD null-K\"ahler metrics are homogeneous of Bianchi type $VIII$.
These Bianchi geometries will be characterised elsewhere.

\section{Acknowledgements}
We thank David Calderbank for drawing our attention to  
the generalised Nahm's equations, and the anonymous referee for
valuable comments. MD was partly supported by NATO grant PST.CLG.978984.


\end{document}